\documentclass[aps,prb,twocolumn,superscriptaddress]{revtex4}
\usepackage{graphicx}
\usepackage{epsfig}
\usepackage{float}

\newcommand{\be}{\begin{equation}}
\newcommand{\ee}{\end{equation}}
\begin{document}

\title{Why the magnetic interactions in Na$_{x}$CoO$_{2}$ are 3D}
\author{M.D. Johannes}
\author{I.I. Mazin}
\affiliation{Code 6391, Naval Research Laboratory, Washington, D.C. 20375}
\author{D.J. Singh}
\affiliation{Condensed Matter Sciences Division, Oak Ridge National
Laboratory, Oak Ridge, TN 37831-6032}

\begin{abstract} 
The puzzle of 3D magnetic interactions in the structurally
2D layered oxide Na$_{x}$CoO$_2$ is addressed using first principles
calculations and analysis of the exchange mechanisms.
The calculations agree with recent neutron
results, favoring AFM stacking of FM planes.
Superexchange via
direct O-O hopping and through intermediate Na $sp^2$ hybrids
couples each Co to its nearest and
six \textit{next}-nearest interplanar neighbors equally.
The individual exchange constants are rather 2D,
like the lattice itself,
but due to multiple c-axis exchange paths,
the magnetism becomes effectively 3D.
\end{abstract}

\maketitle

The layered transition metal oxide (TMO),
Na$_{x}$CoO$_{2}$ is attracting considerable
interest because of
its unusual magnetic and transport properties, and
the recently discovered
superconductivity in its derivative,
Na$_{1/3}$CoO$_{2}\cdot $4/3H$_{2}$O.
Experimental signatures of triplet pairing \cite{WHKO+03,AKAK+03,TWCM+} have
lead to suggestions that spin
fluctuations \cite{DJS03,MDJ+b,KIYI+,ATXH03,TFGZ+}
mediate the superconductivity, in analogy with other layered superconducting
TMO's like high-T$_c$ cuprates and Sr$_2$RuO$_4$,
both of which are close to magnetic instabilities, as is
Na$_{x}$CoO$_{2}$. \cite{DJS00,DJS03}
However, in contrast to those materials
where not only the electronic structure,
but also the magnetic interactions are strongly 2D,
recent neutron measurements of the magnon
dispersion \cite{SPB+04b,LMH+04} in Na$_{x\simeq 0.8}$CoO$_{2}$
indicate that the antiferromagnetic (AFM) exchange
between CoO$_2$ planes is nearly as
strong as the ferromagnetic (FM) in-plane exchange.
Since dimensionality plays a key role in unconventional superconductivity,
it is important to understand
the microscopic physics behind the apparent 3D character
of magnetic interactions observed in Refs.
\onlinecite{SPB+04b,LMH+04} and to clarify the
relationship of the magnetic properties of the parent compound
to those of the superconducting hydrate.  We
address these issues here.

Na$_{x}$CoO$_{2}$ has an unusual magnetic phase diagram as
a function of $x$, which itself poses interesting,
unresolved questions. For most $x$,
Na$_{x}$CoO$_{2}$ is a metallic paramagnet,
but in a very small range, around $x$=0.5, a charge-ordered,
insulating, and possibly
antiferromagnetic (AFM) region emerges \cite{MLF+04}.
Surprisingly, the metallic states on either side of this
region are quite different. For $x < 0.5$
the susceptibility is Pauli paramagnet like with weak $T$ dependence,
while for $x > 0.5$ it is Curie-Weiss like, suggesting local moments.
Finally, for $x$ greater than $\sim 0.75$ a spin density wave condenses,
\cite{TMRU+03,JSHI+03,BCS+04,JWDMP+04}
with clear antiferromagnetism, $T_N \sim 22K$ at $x$=0.82.
While this and the negative Weiss constant \cite{FCC04, JLG+03,YWNSR+03,TMRU+03}
suggest antiferromagnetic interactions,
experiments also 
show a characteristically ferromagnetic
hysteresis \cite{TMRU+03} as well as predicted \cite{DJS03} in-plane
ferromagnetic fluctuations\cite{ATB+03}.

These varied data are reconciled by neutron
scattering experiments, \cite{SPB+04b,LMH+04} which attribute the
$T_{M}$=22K transition to an A-type AFM ordering
(FM planes stacked antiferromagnetically along the $c$-axis).
By fitting magnon
dispersion curves to a linear spin wave model,
both groups conclude that in-plane and perpendicular
magnetic exchange constants are of
comparable magnitude, indicating magnetic isotropy,
despite considerable structural two-dimensionality.
We will show here that this
quasi-isotropy does not imply comparable magnitudes of the
nearest neighbor exchange constants in- and out-of-plane, but rather
unexpectedly large coupling to next nearest neighbors across the planes,
assisted mainly by Na $sp^2$ hybrid orbitals.

Our first principles local density
approximation (LDA) calculations were done using the augmented
plane wave plus local orbital (APW+lo) and linearized augmented
planewave (LAPW) methods implemented in two codes.
\cite{Lnote,Wien2k}.
The experimental lattice parameters,
$a$=2.828\AA,  $c$=10.94\AA, and LDA relaxed
oxygen height $z_{\rm O}$=0.0859
were used for all calculations.\cite{Onote}
The partial occupation of Na was treated both by the
virtual crystal approximation (VCA) and in supercells.
In the VCA, each $2d$ Wyckoff position
of the $P6_3/mmc$ cell was occupied by a fictitious atom, atomic number,
$Z=10+x$, to model a partial occupancy of $x$. We also did some
calculations with Na at the 2$b$ positions.
Supercell calculations were done at selected $x$
with real Na at some $2d$ or $2b$ positions.

Prior calculations show that there is a FM solution in
the LDA \cite{DJS00} for Na$_{x}$CoO$_{2}$ for all $x$ in the
experimentally relevant range. Our
calculations for a tripled $x=2/3$ supercells
show that the A-type AFM order
recently revealed by neutrons is actually the preferred LDA
ground state.
AFM stacking of the ferromagnetic layers is favored by 2.3 meV/Co for
Na in the $2b$ site (Na on top of Co) in a $\sqrt{3}\times\sqrt{3}$ supercell,
and 1.7 meV/Co in the $2d$ site.
This is consistent with a recently measured 
\cite{JLL04} metamagnetic AFM to
FM transition at the relatively low field of 8T.
We find very similar results in the VCA:
1.4 meV/Co with Na in site $2d$ and 2.2 meV/Co with Na in site
$2b$. The total spin moment inside the Co APW spheres for
the A-type AFM ordering is only
$\sim $ 0.02 $\mu _{B}$ less than that in
the FM case which is half metallic.
In other words, the Co is maximally spin polarized for the doping level $x$
consistent with the $\sim$ 0.2 $\mu _{B} $ limit
from neutron \cite{SPB+04} and
muon spin relaxation \cite{JSHI+03} experiments
at $x \sim$=0.8.
However, note that at $x=2/3$, samples are paramagnetic; the LDA ordered
ground state is presumably suppressed by quantum critical fluctuations that
are potentially important for superconductivity. \cite{DJS03}

% The difference between the VCA and the supercell
% calculations is an indication that the two terms are nearly in balance.

We now turn to the relative magnitude of in-plane and
perpendicular exchange constants.
Since the band structure is metallic, antiferromagnetic superexchange
competes with ferromagnetic double exchange
(the former depends on hopping linearly, and the latter quadratically).
The large
in-plane dispersion leads to a net FM in-plane interaction,
while the net inter-plane coupling, due to the smaller $c$-axis dispersion, is
AFM.
As mentioned, nearly isotropic 3D magnetic interactions
are unexpected in layered compounds.
Still this is not inconsistent considering the large bonding-antibonding
splitting of the $a_{1g}$ band in LDA calculations \cite{MDJ+}
at large $x$.
At $x$=2/3, this splitting at $\Gamma$ is 0.21 eV, i.e. 15\%
of the full t$_{2g}$ bandwidth. If the band structure is mapped onto
an effective Co-only model and only nearest neighbor hopping
across the planes is allowed, as assumed in Refs. \onlinecite{SPB+04b}
and \onlinecite{LMH+04} (we will argue that this is
\textit{not} a good approximation), and the effective hopping amplitude
is $t_{\perp }$, then the $a_{1g}$ splitting at $\Gamma$ would be
12$t_{\perp }$ and there would
be no first order splitting for the two $e_{g}^{\prime }$ bands
(consistent with the LDA band structure).
This gives $t_{\perp }\approx 15$ meV.
Although the $t_{2g}-t_{2g}$ in-plane hopping is
nearly an order of magnitude larger,
this still gives less anisotropy than would be anticipated
for a layered material.

To proceed, we need to understand the physics of the interlayer
coupling.
The hopping must proceed,
via O $p$ states, which,
in turn, requires either direct O $p_{z}-p_{z}$ hopping
or hopping assisted by diffuse unoccupied Na
$s$ and $p$ orbitals.
In the latter case the details of Na placement may be important.
Moreover, it is significant that the energy
separation between the unoccupied
Na $3s$ and $3p$ states in Na$_{x}$CoO$_{2}$ is found to be rather small,
smaller in fact than the
corresponding band width.
This allows the $s$ and $p_{x,y}$ orbitals of a Na atom,
sitting inside an O$_{6}$ prism, to combine and form
bands that may be described as coming from $sp^{2}$ hybrid orbitals,
specifically, $h_{1}=(s-\sqrt{2}p_{y})/\sqrt{3},$ $h_{2,3}=(s\pm
\sqrt{3/2}p_{x}+p_{y}/\sqrt{2})/\sqrt{3}.$
These are asymmetric, orthonormal, and directed to
the midpoints of the O prism edges.
So the Na-assisted part of the O-O hopping goes
from one O to another O above it via one, two, or three
Na $sp^{2}$ hybrids
depending on how many of the three nearby Na sites are occupied.
Note that if only Na $s$ states were involved, the hopping
amplitude to a
second nearest neighbor O
in the next plane via a specific Na atom would be the
same as that of hopping to the O right on top. Thus Na $p$
participation changes the interlayer coupling
in an essential way.

\begin{figure} \includegraphics[width=.30\linewidth]{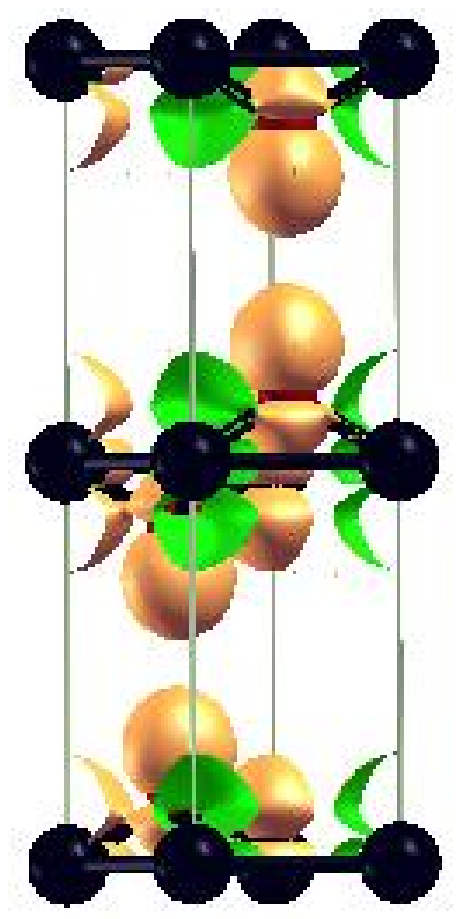}
\includegraphics[width=.316\linewidth]{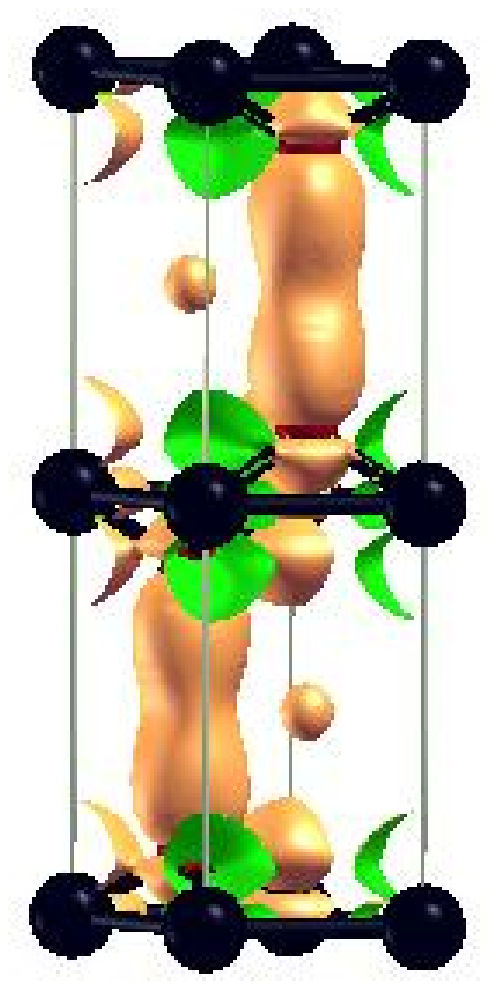}
\includegraphics[width=.27\linewidth]{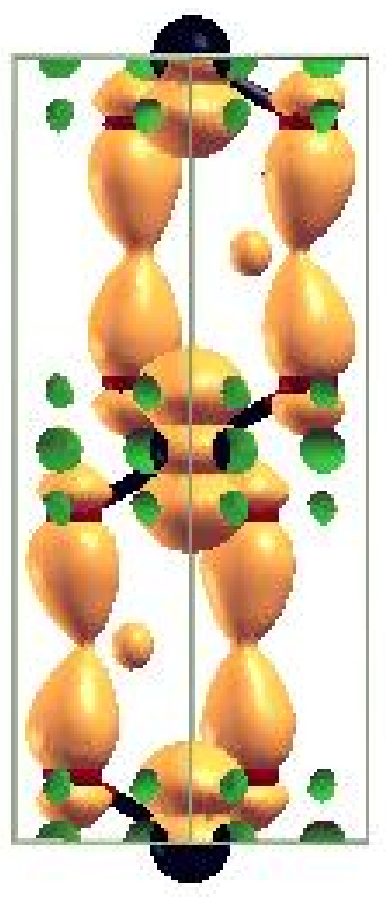} \caption{ (color online) 
Charge density of the
a$_{1g}$ bonding orbital in
X$_{2/3}$CoO$_2$ with X= $e^-$ (left), X= VCA Na (middle),
and X= real Na (right).  The Na ions facilitate O-O bonding and
thus superexchange.
The VCA and supercell calculations differ in the asymmetry
of the $sp^2$ hybrid.} \label{charge}
 \end{figure}

Besides the Na assisted hopping, direct O-O hopping may
be important. Considering the rhombohedral
O site symmetry,
the relevant hopping is from Co $a_{1g}$ to O $p_z$.
Further, in Na$_x$CoO$_2$, the O atoms in
adjacent CoO$_2$ sheets are directly on top of each other,
favoring inter-layer hopping via the opposing O $p_z$ orbitals.
To estimate this
contribution we suppressed the Na-assisted hopping in three ways.
First, we looked at CoO$_{2}+x$ $e^{-}$ i.e., a
compound without Na, but with the correct number of valence
electrons compensated by a uniform background charge. At
$x=2/3,$ the resulting $a_{1g}$ bandsplitting at $\Gamma $ is 0.12 eV
or $\approx $70\% of the total effective coupling in
Na$_{2/3}$CoO$_{2}$.
We verified that this is independent of the amount of extra valence
charge in the system, by doing
calculations with 0.82, 0.75, and 0.4 extra $e^{-}$. Next, for $x=2/3$,
we put Ne in the Na sites, again adding electrons with a
compensating background.
Finally, we did calculations for
$\widetilde{\mathrm{Na}}_{2/3}$CoO$_{2},$
where $\widetilde{\mathrm{Na}}$ denotes Na with $s$ and $p$ orbitals
artificially shifted to higher energy. The results change
very little, indicating that the splitting from direct
O-O hopping is $\sim$ 0.12 eV.
Thus, the effective interplanar
coupling can be written as $t_{\perp }=t_{\perp }^{Na}(x)+t_{\perp }^{O}$,
where $t_{\perp}^{Na}(x)$ is a function of doping and
$t_{\perp }^{O}$ is constant, and $t_{\perp }^{Na}(x)/t_{\perp }^{O}$ varies
from 0.48 to 0.78 in the range $0.6\alt x\alt0.9$.

\begin{figure}[tbp]
\includegraphics[width= 0.9 \linewidth]{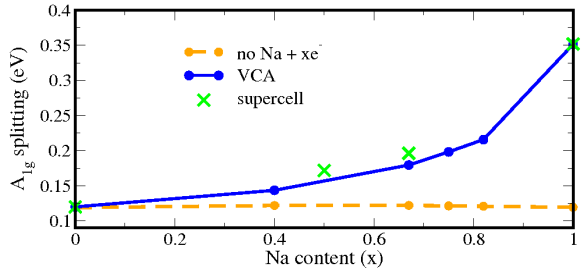}
\caption{A comparison of VCA and supercell a$_{1g}$ splitting at
the $\Gamma$ point, with the same structural parameters(see text), 
showing a monotonic increase with $x$. Calculations without Na orbitals, but with the proper valence charge show that the O-O hopping 
contribution is independent of $x$.}
\label{na}
\end{figure}

The Na assisted hopping can be assessed using VCA or supercell calculations.
These differ in two ways. First, in the VCA, all Na
sites are occupied regardless of $x$, while in a supercell
and in reality
a portion, $1-x$ are empty.
The number of effective hopping paths via
Na, and, correspondingly,
the effective hopping between the planes,
should therefore be reduced in this proportion relative to the
VCA.
Second, the VCA nuclear charge is $Z=10+x$, so the
unoccupied Na $s$ and $p$ energies are moved up.
This will reduce the Na-O hybridization,
which is inversely proportional to the energy separation
between Na 3$s(p)$ and O 2$p$ bands.
This implies a artificial monotonic increase of Na-assisted hopping
with $x$ in the VCA.
These two effects are in opposite directions and should at least
partially cancel.
To assess this,
we performed VCA calculations for $0\leq x\leq 1$, and supercell
(real Na) calculations for $x$=0, 1/2, 2/3 and 1.
The calculated VCA $a_{1g}$ splitting at $\Gamma $ depends on $x$ non-linearly
in both cases; the supercell calculations give similar results with slightly
larger splittings.
(Fig. \ref{na}). We verified that the deviation from
this linearity for the supercell calculations
has the same origin as in the VCA:
the Na 3$s$ and 3$p$ levels shift with $x$
though to a lesser degree than in the VCA (the shift is
due to the changing Coulomb potential as a function of doping).
This is seen in the charge density of the
bonding $a_{1g}$ state at $\Gamma $,
(Fig. \ref{charge}).
The relatively weak O $p_{z}-p_{z}$ overlap
is augmented in the VCA by a
composite state in the middle of the O-O bond,
consisting of the three Na sp$^{2}$ hybrids;
removal of one of these with the
corresponding Na atom in the Na$_{2/3}$CoO$_{2}$
supercell makes the O orbitals tilt towards the remaining Na ions.

A somewhat counterintuitive consequence of the above analysis
is the fact that the nearest neighbor
approximation is invalid for inter-plane hopping.
Neither a Co-O-Na-O-Co path nor a Co-O-O-Co path needs
to end on the Co site directly above the one where it started.
There are 9 different Na-containing paths
connecting nearest neighbor Co ions in different
planes\cite{notepath}, and 3 connecting one Co with each
of the 6 second neighbors\cite{note2bd}.
For paths without Na, there are 3 that connect nearest neighbors
and 1 that connects second neighbors.
Assuming that each Na-containing path contributes $\tau$ to the
effective interplanar hopping amplitude
and each O-only path contributes $\tau ^{\prime}$, we 
find that $t_{\perp }^{Na}(x,\mathbf{k})/t_{\perp }^{O}(\mathbf{k})=3\tau
x/\tau ^\prime$ and
\begin{equation}
t_{\perp }^{O}(\mathbf{k})=\tau ^{\prime }(3+2\cos \mathbf{ak+}2\cos \mathbf{%
bk+}2\cos \mathbf{ck}),
\end{equation}
where \textbf{a}, \textbf{b,} and \textbf{c} are respectively
the three projections onto
the $ab$ plane of the
vectors connecting a given Co ion with the three O above.
At $\Gamma ,$ $t_{\perp }^{Na}=27\tau x$, and $t_{\perp } ^{O}$=0.12
eV, thus we find, at $x$=0.82, $\tau=0.5$ meV and $\tau^{\prime}=1.1$
meV. Finally, we use the number of $\tau $ and $\tau ^{\prime }$ paths that
connect a Co ion to each type of neighbor to calculate $t_{c }$ and $
t_{c}^{\prime }$, the hopping integrals for first and second Co
interplanar coupling to get:
\begin{equation}
\frac{J_{c}}{J_{c}'} = \frac{t_{c}^2}{t_{c}'^2} = \left(\frac{9 \tau
\cdot x
+ 3 \tau '}{3 \tau \cdot x
+ \tau '} \right )^2 = 9
\end{equation}
\begin{figure}[tbp]
\includegraphics[width= 2.1 in, angle=270]{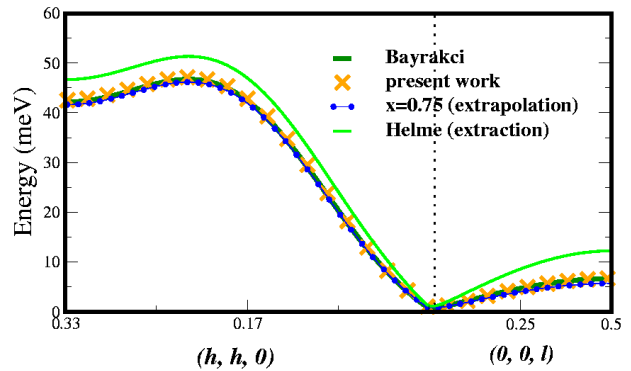}
\caption{Magnon dispersion of Na$_x$CoO$_2$ along high symmetry lines shown
for $x$= 0.7 and $x$=0.82. The results of two experiments in which data were
fit to a nearest-neighbor interaction model are compared to our dispersion
which includes second-nearest out-of-plane neighbors. Our parameters were
fit to the data of Ref. \onlinecite{SPB+04b} for $x$=0.7,
adjusted for Na content (see text), and used to get the
dispersion for $x$=0.82 \textit{without} re-fitting.}
\label{omega}
\end{figure}

We now revisit the interpretation of the spin wave dispersion observed
in Refs. \onlinecite{SPB+04b} and \onlinecite{LMH+04},
including second neighbor assisted hopping. A
straightforward generalization of Eq. 2 in Ref. \onlinecite{SPB+04b} yields 
\[
E =2S\sqrt{\{\tilde{\mathcal{J}}_{\perp }(0)-\mathcal{J}%
_{\parallel }(0)+\mathcal{J}_{\parallel }(\mathbf{q})+ (\frac{D}{2S})\}^{2}-\{\tilde{
\mathcal{J}}_{\perp }(\mathbf{q})\}^{2}}
\]
where, following their notation, $hkl$ is
the wave vector in units of the reciprocal lattice vector. $\tilde{\mathcal{J
}}(\mathbf{q})=\mathcal{J}_{c }(\mathbf{q})+\mathcal{J}_{c}^{\prime
}(\mathbf{q})$ and $\mathcal{J}_{c}^{\prime }(\mathbf{q})=2J_{c
}^{\prime }\cos (\pi l)[2\cos (2\pi h)+2\cos (2\pi k)+2\cos (2\pi (h+k)]$.
Using the $\mathbf{k}=(0,0,1/2)$ magnon energy, we can extract $J_{c
}=1.98$ meV and $J_{c}^{\prime }=$0.22 meV. This fits experiment 
\cite{SPB+04b} well with $J_{\parallel }=-4$.5 meV. Bayracki $et$ $al$,
using nearest neighbors only, found $J_{c}=3.3$ meV and $J_{\parallel
}=-4.5$ meV, leading to the conclusion of magnetic isotropy. Fig. \ref{omega}
shows that our model produces results that are completely indistinguishable
from the nearest neighbor model used in Ref. \onlinecite{SPB+04b}, however,
the parameters are more physically reasonable and compatible with the
highly anisotropic electronic structure of the compound.

Finally, we address the differences in the magnon dispersions
obtained in Ref. \onlinecite{SPB+04b} and Ref. \onlinecite{LMH+04},
both shown in Fig.  \ref{omega}.
Because of notational differences between Ref. \onlinecite{SPB+04b} and Ref.
\onlinecite{LMH+04}, we extract exchange constants
directly from the Ref. \onlinecite{LMH+04} data, using the same formalism as
Ref. \onlinecite{SPB+04b}
and get $J_{c}$ = 6.1 eV and $J_{\parallel }=-4$.5 eV.
Using VCA calculations at $x$=0.7 and $x$=0.82,
we scale the parameters of our nearest and
next-nearest neighbor model and compare them with the results of Helme $et$
$al$.\cite{LMH+04}
We estimate that at $x=0.75$ $J_{c }$=1.69 meV,
$J_{c }^{\prime }=$0.19 meV, and $J_{\parallel }=-$4.5 meV. So,
the in-plane hopping remains unchanged, in agreement with experiment,
and the inter-planar hopping decreases by $ \approx 8\%,$
since less Na is available to mediate it.

The measured dispersion\cite{LMH+04}
in the $z$ direction is much \textit{larger} at $x$=0.75 than at $x$=0.82.
Other effects, not
accounted for here, must therefore be operative.
One possibility is a different pattern of Na ordering.
Indeed, 
in our $x=2/3$ supercell calculations
placing all the Na at the $2b$ sites instead of the $2d$ sites
yields a $\sim$ 35\% increase in the interlayer
coupling. While this number no doubt depends on the exact Na
arrangement, it does indicate sensitivity to the ordering of a size
sufficient to explain the experiment.
In comparing with experiment, it should be also
noted that Co disproportionation,
seen by NMR studies \cite{IRM+04}around $x$=0.7,
would change the spin arrangement of the
lattice and therefore the magnon dispersion.
Co ions with formal valency (4-x)$^{+}$ split proportionally into non-magnetic
Co$^{3+}$ and $S=1/2$ Co$ ^{4+}$.
The particular arrangement is unknown,
but the exchange interaction may be changed.
Note that the phase boundary of charge order is not yet clearly defined
and calculations \cite{JKK-L03} suggest that, near the crossover,
there is a region where two distinct magnetic ions exist.
Another consideration
is that the spin-wave model adopted
here and in Refs.  \onlinecite{SPB+04b} and \onlinecite{LMH+04} assumes
a rigid spin moment of S=1/2,
which may not be a good approximation for the weak magnetism in this compound.
The actual moment,
both measured and calculated,
is smaller and grows with $x$.
Last but not least,
one should keep in mind that FM and AFM
interactions here are competing with each other
thereby amplifying the relative effect of doping changes on magnetism.  

In summary, we find that inter-plane coupling to second neighbors
plays an important role in the out-of-plane magnetism of
Na$_{x}$CoO$_{2}.$ Heisenberg type models including
only one nearest neighbor exchange across the plane are
insufficient;
exchange between the six next nearest neighbors in adjacent planes is needed.
Fortunately, one can estimate the ratio of the two exchange constants,
making it possible to extract them from experiment
without increasing the number of fitting parameters.
The resulting exchange constants provide a magnon spectrum that matches
experiment extremely well and yields a physically realistic
picture of magnetic interactions in this layered material.
The LDA energy difference between the FM
and the A-type AFM ordering for Na content $x=2/3$ is $\sim 2$
meV/Co, which is reasonable considering the exchange coupling deduced
from experimentally measured spin wave dispersion
(6.6 meV and 12.1 meV/Co in
Refs. \onlinecite{SPB+04b} and \onlinecite{LMH+04}, respectively).
We note that superexchange drops with the distance
more strongly than other coupling mechanisms,
such as double exchange. Thus, although the
inter-planar coupling is surprisingly
strong in Na$_x$CoO$_2$,
yielding three dimensional magnetic character, it would be expected
to be very weak in the hydrated superconducting compound.
It is even possible that a cross-over from AFM to FM coupling occurs
with hydration due to suppression of the superexchange interaction.
If the magnetic interactions become effectively 2D in the
hydrated compound, fluctuations would be enhanced,
possibly revealing superconductivity associated with a nearby magnetic quantum
critical point.

We are grateful for helpful discussions with
A.T. Boothroyd, R. Jin and S. Nagler. Work
at Oak Ridge National Laboratory is supported by the U.S. Department of
Energy.


\begin{thebibliography}{31}
\expandafter\ifx\csname natexlab\endcsname\relax\def\natexlab#1{#1}\fi
\expandafter\ifx\csname bibnamefont\endcsname\relax
  \def\bibnamefont#1{#1}\fi
\expandafter\ifx\csname bibfnamefont\endcsname\relax
  \def\bibfnamefont#1{#1}\fi
\expandafter\ifx\csname citenamefont\endcsname\relax
  \def\citenamefont#1{#1}\fi
\expandafter\ifx\csname url\endcsname\relax
  \def\url#1{\texttt{#1}}\fi
\expandafter\ifx\csname urlprefix\endcsname\relax\def\urlprefix{URL }\fi
\providecommand{\bibinfo}[2]{#2}
\providecommand{\eprint}[2][]{\url{#2}}

\bibitem[{\citenamefont{{Y. Wang} et~al.}(2003)\citenamefont{{Y. Wang}, {N. S.
  Rogado}, {R. J. Cava}, and {N. P. Ong}}}]{YWNSR+03}
\bibinfo{author}{\bibnamefont{{Y. Wang \it{et al}}}}, 
\bibinfo{journal}{Nature}
  \textbf{\bibinfo{volume}{423}}, \bibinfo{pages}{425} (\bibinfo{year}{2003}).

\bibitem[{\citenamefont{{W. Higemoto} et~al.}(2004)\citenamefont{{W. Higemoto},
  {K. Ohishi}, {A. Koda}, {R. Kadono}, {K. Ishida}, {K. Takada}, {H. Sakurai},
  {E. Takayama-Muromachi}, and {T. Sasaki}}}]{WHKO+03}
\bibinfo{author}{\bibnamefont{{W. Higemoto \it{et al}}}},
\bibinfo{journal}{Phys. Rev. B}
  \textbf{\bibinfo{volume}{70}}, \bibinfo{pages}{134508}
  (\bibinfo{year}{2004}).

\bibitem[{\citenamefont{{A. Kanigel} et~al.}(2004)\citenamefont{{A. Kanigel},
  {A. Keren}, {L. Patlagan}, {K. B. Chashka}, {P. King}, and {A.
  Amato}}}]{AKAK+03}
\bibinfo{author}{\bibnamefont{{A. Kanigel \it{et al}}}}, 
\bibinfo{journal}{Phys. Rev.
  Lett.} \textbf{\bibinfo{volume}{92}}, \bibinfo{pages}{257007}
  (\bibinfo{year}{2004}).

\bibitem[{\citenamefont{{T. Waki} et~al.}(2003)\citenamefont{{T. Waki}, {C.
  Michioka}, {M. Kato}, {K. Yoshimura}, {K. Takada}, {H. Sakurai}, {E.
  Takayama-Muromachi}, and {T. Takayoshi Sasaki}}}]{TWCM+}
\bibinfo{author}{\bibnamefont{{T. Waki \it{et al}}}}, 
  \bibinfo{journal}{cond-mat/0306036}  (\bibinfo{year}{2003}).

\bibitem[{\citenamefont{{D. J. Singh}}(2003)}]{DJS03}
\bibinfo{author}{\bibnamefont{{D. J. Singh}}}, \bibinfo{journal}{Phys. Rev. B}
  \textbf{\bibinfo{volume}{68}}, \bibinfo{pages}{020503}
  (\bibinfo{year}{2003}).

\bibitem[{\citenamefont{{M. D. Johannes}
  et~al.}(2004{\natexlab{a}})\citenamefont{{M. D. Johannes}, {I. I. Mazin}, {D.
  J. Singh}, and {D. A. Papaconstantopoulos}}}]{MDJ+b}
\bibinfo{author}{\bibnamefont{{M. D. Johannes}}},
  \bibinfo{author}{\bibnamefont{{I. I. Mazin}}},
  \bibinfo{author}{\bibnamefont{{D. J. Singh}}}, \bibnamefont{and}
  \bibinfo{author}{\bibnamefont{{D. A. Papaconstantopoulos}}},
  \bibinfo{journal}{PRL} \textbf{\bibinfo{volume}{93}}, \bibinfo{pages}{097005}
  (\bibinfo{year}{2004}{\natexlab{a}}).

\bibitem[{\citenamefont{{K. Ishida} et~al.}(2003)\citenamefont{{K. Ishida}, {Y.
  Ihara}, {Y. Maeno}, {C. Michioka}, {M. Kato}, {K. Yoshimura}, {K. Takada},
  {T. Sasaki}, {H. Sakurai}, and {E. Takayama-Muromachi}}}]{KIYI+}
\bibinfo{author}{\bibnamefont{{K. Ishida \it{et al}}}}, 
\bibinfo{journal}{J.
  Phys. Soc. Jpn.} \textbf{\bibinfo{volume}{72}}, \bibinfo{pages}{3041}
  (\bibinfo{year}{2003}).

\bibitem[{\citenamefont{{A. Tanaka} and {X. Hu}}(2003)}]{ATXH03}
\bibinfo{author}{\bibnamefont{{A. Tanaka}}} \bibnamefont{and}
  \bibinfo{author}{\bibnamefont{{X. Hu}}}, \bibinfo{journal}{Phys. Rev. Lett.}
  \textbf{\bibinfo{volume}{91}}, \bibinfo{pages}{257006}
  (\bibinfo{year}{2003}).

\bibitem[{\citenamefont{{T. Fujimoto} et~al.}(2004)\citenamefont{{T. Fujimoto},
  {G. Zheng}, {Y. Kitaoka}, {R. L. Meng}, {J. Cmaidalka}, and {C. W.
  Chu}}}]{TFGZ+}
\bibinfo{author}{\bibnamefont{{T. Fujimoto \it{et al}}}},
\bibinfo{journal}{Phys. Rev.
  Lett.} \textbf{\bibinfo{volume}{92}}, \bibinfo{pages}{047004}
  (\bibinfo{year}{2004}).

\bibitem[{\citenamefont{{M. L. Foo} et~al.}(2004)\citenamefont{{M. L. Foo}, {Y.
  Wang}, {S. Watuchi}, {H. W. Zandbergen}, {T. He}, {R. J. Cava}, and {N. P.
  Ong}}}]{MLF+04}
\bibinfo{author}{\bibnamefont{{M. L. Foo \it{et al}}}}, 
  \bibinfo{journal}{Phys. Rev. Lett} \textbf{\bibinfo{volume}{92}},
  \bibinfo{pages}{247001} (\bibinfo{year}{2004}).

\bibitem[{\citenamefont{{T. Motohashi} et~al.}(2003)\citenamefont{{T.
  Motohashi}, {R. Ueda}, {E. Naujalis}, {R. Tojo}, {I. Terasaki}, {T. Atake},
  {M. Karppinen}, and {H. Yamauchi}}}]{TMRU+03}
\bibinfo{author}{\bibnamefont{{T. Motohashi \it{et al}}}},
\bibinfo{journal}{Phys. Rev. B} \textbf{\bibinfo{volume}{67}},
  \bibinfo{pages}{064406} (\bibinfo{year}{2003}).

\bibitem[{\citenamefont{{J. Sugiyama} et~al.}(2003)\citenamefont{{J. Sugiyama},
  {H. Itahara}, {J. H. Brewer}, {E. J. Ansaldo}, {T. Motohashi}, {M.
  Karppinen}, and {H. Yamauchi}}}]{JSHI+03}
\bibinfo{author}{\bibnamefont{{J. Sugiyama \it{et al}}}},
\bibinfo{journal}{Phys. Rev.
  B} \textbf{\bibinfo{volume}{67}}, \bibinfo{pages}{214420}
  (\bibinfo{year}{2003}).

\bibitem[{\citenamefont{{B. C. Sales} et~al.}(2004)\citenamefont{{B. C. Sales},
  {R. Jin}, {K. A. Affholter}, {P. Khalifah}, {G. M. Veith}, and {D.
  Mandrus}}}]{BCS+04}
\bibinfo{author}{\bibnamefont{{B. C. Sales \it{et al}}}},
  \bibinfo{journal}{Phys. Rev. B} \textbf{\bibinfo{volume}{70}}, \bibinfo{pages}{174419} (\bibinfo{year}{2004}).

\bibitem[{\citenamefont{{J. Wooldridge} et~al.}(2004)\citenamefont{{J.
  Wooldridge}, {D. M. Paul}, {G. Balakrishnan}, and {M. R. Lees}}}]{JWDMP+04}
\bibinfo{author}{\bibnamefont{{J. Wooldridge}}},
  \bibinfo{author}{\bibnamefont{{D. M. Paul}}},
  \bibinfo{author}{\bibnamefont{{G. Balakrishnan}}}, \bibnamefont{and}
  \bibinfo{author}{\bibnamefont{{M. R. Lees}}},
  \bibinfo{journal}{cond-mat/0406513}  (\bibinfo{year}{2004}).


\bibitem[{\citenamefont{{J. L. Gavilano} et~al.}(2003)\citenamefont{{J. L.
  Gavilano}, {D. Rau}, {B. Pedrini}, {J. Hinderer}, {H. R. Ott}, {S. Kazakov},
  and {J. Karpinski}}}]{JLG+03}
\bibinfo{author}{\bibnamefont{{J. L. Gavilano \it{et al}}}},
  \bibinfo{journal}{cond-mat/0308383} \textbf{\bibinfo{volume}{69}}, \bibinfo{pages}{100404} (\bibinfo{year}{2004}).

\bibitem[{\citenamefont{{F. C. Chou} et~al.}(2004)\citenamefont{{F. C. Chou},
  {J. H. Cho}, and {Y. S. Lee}}}]{FCC04}
\bibinfo{author}{\bibnamefont{{F. C. Chou}}}, \bibinfo{author}{\bibnamefont{{J.
  H. Cho}}}, \bibnamefont{and} \bibinfo{author}{\bibnamefont{{Y. S. Lee}}},
  \bibinfo{journal}{Phys. Rev. B} \textbf{\bibinfo{volume}{70}},
  \bibinfo{pages}{144526} (\bibinfo{year}{2004}).

\bibitem[{\citenamefont{{A. T. Boothroyd} et~al.}(2003)\citenamefont{{A. T.
  Boothroyd}, {R. Coldea}, {D. A. Tennant}, {D. Prabhakaran}, and {C. D.
  Frost}}}]{ATB+03}
\bibinfo{author}{\bibnamefont{{A. T. Boothroyd \it{et al}}}},
\bibinfo{journal}{Phys. Rev.
  Lett.} \textbf{\bibinfo{volume}{92}}, \bibinfo{pages}{197201}
  (\bibinfo{year}{2003}).

\bibitem[{\citenamefont{{D. J. Singh}}(2000)}]{DJS00}
\bibinfo{author}{\bibnamefont{{D. J. Singh}}}, \bibinfo{journal}{Phys. Rev. B}
  \textbf{\bibinfo{volume}{61}}, \bibinfo{pages}{13397} (\bibinfo{year}{2000}).

\bibitem[{\citenamefont{{S. P. Bayrakci}
  et~al.}(2004{\natexlab{a}})\citenamefont{{S. P. Bayrakci}, {I. Mirebeau}, {P.
  Bourges}, {Y. Sidis}, {M. Enderle}, {J. Mesot}, {D. P. Chen}, {C. T. Lin},
  and {B. Keimer}}}]{SPB+04b}
\bibinfo{author}{\bibnamefont{{S. P. Bayrakci \it{et al}}}},
  \bibinfo{journal}{cond-mat/0410224}  (\bibinfo{year}{2004}{\natexlab{a}}).

\bibitem[{\citenamefont{{L. M. Helme} et~al.}(2004)\citenamefont{{L. M. Helme},
  {A. T. Boothroyd}, {R. Coldea}, {D. Prabhakaran}, {D. A. Tennant}, {A.
  Hiess}, and {J. Kulda}}}]{LMH+04}
\bibinfo{author}{\bibnamefont{{L. M. Helme \it{et al}}}},
  \bibinfo{journal}{cond-mat/0410457}  (\bibinfo{year}{2004}).

\bibitem[{Lno()}]{Lnote}
\bibinfo{note}{We did well converged
calculations with the Wien2k code \cite{Wien2k}
and an independent LAPW code. These gave practically identical
results when tested for the same system.
For the APW+lo calculations sphere radii of
1.9 $a_0$,
1.6 $a_0$ and 2.0 $a_0$ were used for Co, O and Na, respectively.
For the LAPW calculations the corresponding radii were
1.95 $a_0$, 1.55 $a_O$ and 2.0 $a_O$.
}

\bibitem[{\citenamefont{{P. Blaha} et~al.}(2002)\citenamefont{{P. Blaha}, {K.
  Schwarz}, {G. K. H. Madsen}, {D. Kvasnicka}, and {J. Luitz}}}]{Wien2k}
\bibinfo{author}{\bibnamefont{{P. Blaha}}}, \bibinfo{author}{\bibnamefont{{K.
  Schwarz}}}, \bibinfo{author}{\bibnamefont{{G. K. H. Madsen}}},
  \bibinfo{author}{\bibnamefont{{D. Kvasnicka}}}, \bibnamefont{and}
  \bibinfo{author}{\bibnamefont{{J. Luitz}}}, \emph{\bibinfo{title}{Wien2k}}
  (\bibinfo{year}{2002}), \bibinfo{note}{iSBN 3-9501031-1-2}.

% \bibitem[{\citenamefont{{J. P. Perdew} and {Y. Wang}}(1992)}]{LDA5}
% \bibinfo{author}{\bibnamefont{{J. P. Perdew}}} \bibnamefont{and}
% \bibinfo{author}{\bibnamefont{{Y. Wang}}}, \bibinfo{journal}{Phys. Rev. B}
% \textbf{\bibinfo{volume}{45}}, \bibinfo{pages}{13244} (\bibinfo{year}{1992}).

\bibitem[{Ono()}]{Onote}
\bibinfo{note}{We verified that relaxing the O position as a function of $x$
has relatively little effect, and keeping the position fixed helps elucidate
  the physics related to Na doping.}

\bibitem[{\citenamefont{{J. L. Luo} et~al.}(2004)\citenamefont{{J. L. Luo},
  {N.L. Wang}, { G.T. Liu}, {D. Wu}, {X.N. Jing}, {F. Hu}, and {T.
  Xiang}}}]{JLL04}
\bibinfo{author}{\bibnamefont{{J. L. Luo \it{et al}}}},
\bibinfo{journal}{Phys. Rev.
  Lett} \textbf{\bibinfo{volume}{93}}, \bibinfo{pages}{187203}
  (\bibinfo{year}{2004}).

\bibitem[{\citenamefont{{S. P. Bayrakci}
  et~al.}(2004{\natexlab{b}})\citenamefont{{S. P. Bayrakci}, {C. Bernhard}, {D.
  P. Chen}, {B. Keimer}, {R. K. Kermer}, {P. Lemmens}, {C. T. Lin}, {C.
  Niedermayer}, and {J. Strempfer}}}]{SPB+04}
\bibinfo{author}{\bibnamefont{{S. P. Bayrakci \it{et al}}}},
\bibinfo{journal}{Phys. Rev.
  B} \textbf{\bibinfo{volume}{69}}, \bibinfo{pages}{100410}
  (\bibinfo{year}{2004}{\natexlab{b}}).

\bibitem[{\citenamefont{{M. D. Johannes}
  et~al.}(2004{\natexlab{b}})\citenamefont{{M. D. Johannes}, {D. A.
  Papaconstantopoulos}, {D. J. Singh}, and {M. J. Mehl}}}]{MDJ+}
\bibinfo{author}{\bibnamefont{{M. D. Johannes}}},
  \bibinfo{author}{\bibnamefont{{D. A. Papaconstantopoulos}}},
  \bibinfo{author}{\bibnamefont{{D. J. Singh}}}, \bibnamefont{and}
  \bibinfo{author}{\bibnamefont{{M. J. Mehl}}}, \bibinfo{journal}{Euro. Phys.
  Lett.} \textbf{\bibinfo{volume}{68}}, \bibinfo{pages}{433}
  (\bibinfo{year}{2004}{\natexlab{b}}).

\bibitem[{not({\natexlab{a}})}]{notepath}
\bibinfo{note}{Note that if hopping was only via Na s orbitals, and not Na
  sp$^{2}$ hybrids, this counting would be different.}

\bibitem[{not({\natexlab{b}})}]{note2bd}
\bibinfo{note}{This is independent of whether the Na is at the $2b$
  or the $2d$ sites.}

\bibitem[{\citenamefont{{I. R. Mukhamedshin} et~al.}(2004)\citenamefont{{I. R.
  Mukhamedshin}, {H. Alloul}, {G. Collin}, and {N. Blanchard}}}]{IRM+04}
\bibinfo{author}{\bibnamefont{{I. R. Mukhamedshin}}},
  \bibinfo{author}{\bibnamefont{{H. Alloul}}},
  \bibinfo{author}{\bibnamefont{{G. Collin}}}, \bibnamefont{and}
  \bibinfo{author}{\bibnamefont{{N. Blanchard}}}, \bibinfo{journal}{Phys. Rev.
  Lett.} \textbf{\bibinfo{volume}{93}}, \bibinfo{pages}{167601}
  (\bibinfo{year}{2004}).

\bibitem[{\citenamefont{{K. -W. Lee} et~al.}(2004)\citenamefont{{K. -W. Lee},
  {J. Kunes}, and {W. E. Pickett}}}]{JKK-L03}
\bibinfo{author}{\bibnamefont{{K. -W. Lee}}}, \bibinfo{author}{\bibnamefont{{J.
  Kunes}}}, \bibnamefont{and} \bibinfo{author}{\bibnamefont{{W. E. Pickett}}},
  \bibinfo{journal}{Phys. Rev. B} \textbf{\bibinfo{volume}{70}},
  \bibinfo{pages}{045104} (\bibinfo{year}{2004}).

\end{thebibliography}
\end{document}